\documentclass[sigconf]{acmart}
\AtBeginDocument{%
  \providecommand\BibTeX{{%
    \normalfont B\kern-0.5em{\scshape i\kern-0.25em b}\kern-0.8em\TeX}}}

\copyrightyear{2026}
\acmYear{2026}
\setcopyright{cc}
\setcctype{by}
\acmConference[CIKM '26]{Proceedings of the 35th ACM International Conference on Information and Knowledge Management}{November 07--11, 2026}{Rome, Italy}
\acmBooktitle{Proceedings of the 35th ACM International Conference on Information and Knowledge Management (CIKM '26), November 07--11, 2026, Rome, Italy}
\acmDOI{10.1145/3799682.3841035}
\acmISBN{979-8-4007-2539-5/2026/11}

\usepackage{siunitx}  
\usepackage{graphics}
\usepackage{mdwlist}
\usepackage{balance}  

\usepackage{IEEEtrantools}
\usepackage{amsmath,amsthm,amsfonts,bm}
\usepackage{pifont}

\usepackage{graphicx}
\usepackage{caption}
\usepackage{subcaption}

\usepackage{array}
\usepackage{booktabs}
\usepackage{makecell}
\usepackage{multirow}
\usepackage{multicol}
\usepackage{hyperref}
\usepackage{url}
\usepackage{enumitem}
\usepackage{colortbl}
\usepackage{xcolor}

\newcommand{\cy}{\cellcolor{yellow!30}}
\newcommand{\cl}{\cellcolor{red!20}}
\newcommand{\cd}{\cellcolor{red!45}}
\newcolumntype{C}{>{\centering\arraybackslash}p{1cm}}

\begin{document}

\newcommand{\goal}[1]{ {\noindent {$\Rightarrow$} \em {#1} } }


\title[ELASTIC: Trajectory-Based Synchronization of Event and Tracking Data in Soccer]{ELASTIC: Trajectory-Based Synchronization of \\ Event and Tracking Data in Soccer}



\author{Hyunsung Kim}
\affiliation{
  \institution{KAIST}
  \city{Daejeon}
  \country{Republic of Korea}
}
\additionalaffiliation{%
  \institution{Fitogether Inc.}
  \city{Seoul}
  \country{Republic of Korea}
}
\email{hyunsung.kim@kaist.ac.kr}

\author{Hoyoung Choi}
\affiliation{
    \institution{KAIST}
    \city{Daejeon}
    \country{Republic of Korea}
}
\email{chy3724@kaist.ac.kr}

\author{Kunhee Lee}
\affiliation{
    \institution{KAIST}
    \city{Daejeon}
    \country{Republic of Korea}
}
\email{kunhee8@kaist.ac.kr}

\author{Sangwoo Seo}
\affiliation{
    \institution{KAIST}
    \city{Daejeon}
    \country{Republic of Korea}
}
\email{sangwooseo@kaist.ac.kr}

\author{Tom Boomstra}
\affiliation{
    \institution{AFC Ajax}
    \city{Amsterdam}
    \country{Netherlands}
}
\email{t.boomstra@ajax.nl}

\author{Jinsung Yoon}
\affiliation{
    \institution{Fitogether Inc.}
    \city{Seoul}
    \country{Republic of Korea}
}
\email{jinsung.yoon@fitogether.com}

\author{Chanyoung Park}
\affiliation{
    \institution{KAIST}
    \city{Daejeon}
    \country{Republic of Korea}
}
\email{cy.park@kaist.ac.kr}

\renewcommand{\shortauthors}{Hyunsung Kim et al.}
    
\begin{abstract}
	Combining event and tracking data is fundamental to modern soccer analytics, yet the two sources are rarely well aligned: event timestamps recorded by human annotators often miss the true moment of the action, distorting the spatiotemporal context that downstream models rely on. Existing synchronization methods depend on noisy human-annotated event locations and fail to detect ball receptions, obscuring when each player gains ball possession. To address these limitations, we propose ELASTIC (\textbf{E}vent-\textbf{L}ocation-\textbf{A}gno\textbf{STIC} synchronizer), a framework that infers the start and end timestamps of events solely from player and ball trajectories, without relying on annotated event locations. To recover ball receptions, ELASTIC enriches the event sequence by inserting virtual termination events between consecutive events, so that the end of each event is detected jointly with its start. It then extracts a sparse set of candidate frames where ball touches are physically plausible, and aligns the termination-inserted event sequence with the candidate-frame sequence using an extended Needleman-Wunsch algorithm. For reproducible evaluation, we construct a publicly available benchmark by annotating ground-truth timestamps on the Sportec Open DFL Dataset, on which ELASTIC substantially outperforms existing methods. Through downstream task evaluation, we further show that improved synchronization translates into measurable gains in soccer analytics. The source code and benchmark are available at \url{https://github.com/hyunsungkim-ds/elastic.git}.
\end{abstract}

\ccsdesc[500]{Information systems~Data mining}
\ccsdesc[500]{Information systems~Information integration}
\ccsdesc[300]{Information systems~Spatial-temporal systems}
\ccsdesc[300]{Theory of computation~Dynamic programming}

\keywords{Sports Analytics; Data Quality; Multimodal Data Fusion; Time-Series Synchronization; Sequence Alignment}


\maketitle

\section{Introduction}
\label{sec:intro}
The increasing availability of fine-grained in-game data has reshaped the landscape of soccer analytics. Among the various data modalities, two stand out as primary pillars: \textit{event data} and \textit{tracking data}. The former consists of manually annotated records of key on-ball actions such as passes, shots, and tackles, whereas the latter captures the positions of all players and the ball at every moment of the game through optical or wearable sensor-based systems. When properly combined, these two sources support a wide range of downstream tasks, such as quantifying the scoring probability of a shot~\cite{anzer2021goal,davis2024biases,lucey2015quality,scholtes2024bayesxg,mead2023expected}, analyzing passing options in a given situation \cite{anzer2022expected,fernandez2020soccermap,fernandez2021framework,power2017not,rahimian2022penetrate,rahimian2023pass,spearman2017physicsbased}, and evaluating players' abilities~\cite{everett2025evaluating,kim2026better,robberechts2023unxpass,stockl2021making,teranishi2023evaluation}.

In practice, however, aligning these two sources is challenging. Event data is typically annotated by humans, meaning that the recorded timestamp often does not precisely correspond to the moment when the player actually performed the action. When such event records are naively combined with tracking data, the resulting player and ball configurations may be inaccurate, thereby distorting the spatiotemporal context used in downstream tasks. For this reason, synchronizing event and tracking data has become a central problem in soccer analytics~\cite{biermann2023synchronization,vanroy2023etsy,oonk2025right}. More generally, the task belongs to a broader class of multimodal sequence alignment problems, in which sparse semantic records (i.e., event data) must be precisely aligned with dense temporal signals (i.e., tracking data).

\begin{figure}[t]
\centering
\subfloat[Snapshot\label{fig:sample_snapshot}]{
	\includegraphics[width=0.20\textwidth]{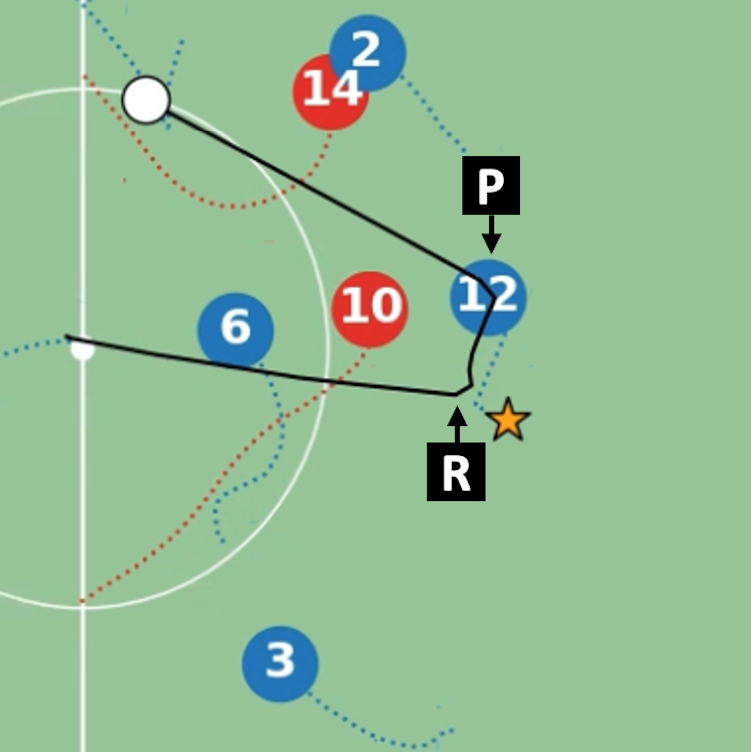}
}
\subfloat[Event-ball distance\label{fig:sample_feats}]{
	\includegraphics[width=0.25\textwidth]{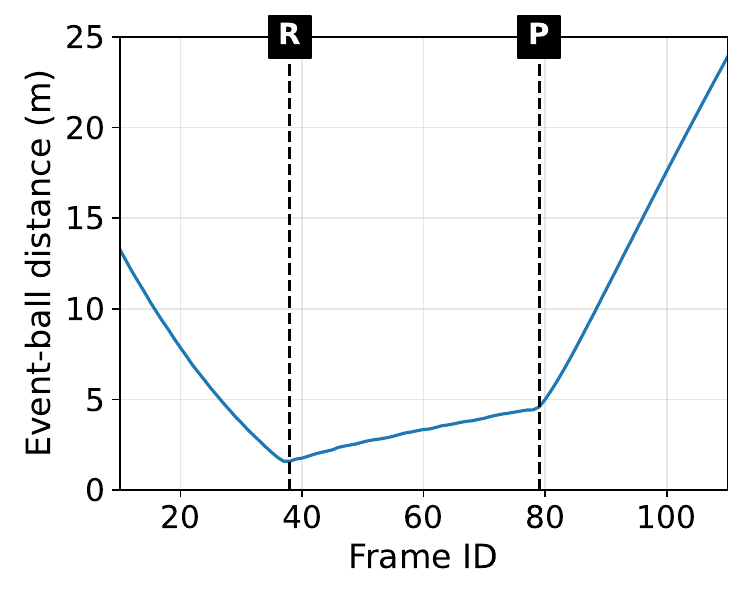}
}
\vspace{-0.5em}
\caption{A sample pass by player~\#12, where R and P mark the locations and moments of the ball reception (frame 38) and the pass (frame 79), respectively. (a) The white circle with a black tail shows the ball trajectory, and the star marks the annotated pass location, which deviates from the ball's actual position at P. (b) The distance between the ball and the annotated pass location over time reaches its minimum near R rather than P, misleading distance-based synchronizers.}
\vspace{-0.5em}
\label{fig:sample}
\end{figure}

To address this issue, several methods have been proposed to synchronize soccer event and tracking data. Biermann et al.~\cite{biermann2023synchronization} introduced a learning-based approach that applies a sliding window over tracking data and uses aggregated window features to classify whether each window's center frame corresponds to a pass. Anzer et al.~\cite{anzer2021goal} and Van Roy et al.~\cite{vanroy2023etsy} proposed distance-based methods that define a window around the annotated event timestamp and find the frame that minimizes the sum of distances between the annotated event location, ball location, and event player location. Oonk et al.~\cite{oonk2025right,oonk2026databallpy} used the Needleman-Wunsch algorithm~\cite{needleman1970general} to find the globally optimal alignment between the two modalities while preserving the order of events. Meanwhile, another line of work detects events directly from tracking data without using event data at all~\cite{vidalcodina2022automatic, bischofberger2024event, mills2026automatic}, but such approaches recover only a limited set of event types defined by their own detection rules. We therefore focus on synchronizing the fine-grained event records that already exist, rather than regenerating them.

Existing synchronization methods, however, share several limitations. First, they depend on manual annotations that are noisy and labor-intensive. Anzer et al.~\cite{anzer2021goal}, Van Roy et al.~\cite{vanroy2023etsy}, and Oonk et al.~\cite{oonk2025right,oonk2026databallpy} use the distance between the human-annotated event location and the ball as a primary scoring criterion, favoring frames with small event-ball distances. Since the annotated event locations are themselves spatially noisy, this criterion can select an incorrect frame. Fig.~\ref{fig:sample} illustrates such a failure: player \#12 receives the ball at frame 38 (R) and passes it at frame 79 (P), but the annotated pass location (the star) lies closer to the reception point (R) than to the position where the pass is actually made (P), causing these methods to wrongly select the receiving moment as the estimated pass timestamp. Biermann et al.~\cite{biermann2023synchronization} instead rely solely on tracking features, but their method requires ground-truth labels to learn the feature distributions of pass and non-pass windows, and operates on coarse window-level features rather than fine frame-level cues, which limits its final accuracy.

In addition, existing methods do not detect ball-receiving events. Anzer et al.~\cite{anzer2021goal} synchronize only shots, Biermann et al.~\cite{biermann2023synchronization} focus only on passes, and Oonk et al.~\cite{oonk2025right,oonk2026databallpy} extend coverage to passes, shots, tackles, and dribbles, but none of them detects the moment when a player receives the ball. This is partly because many data providers do not explicitly record ball-receiving events; as a result, even Van Roy et al.~\cite{vanroy2023etsy}, whose method covers all recorded events, leave them unsynchronized. However, the absence of accurate ball-reception timing precludes downstream analyses such as reconstructing each player's ball possession intervals~\cite{link2017individual} or analyzing passes conditioned on their destinations~\cite{spearman2017physicsbased, fernandez2020soccermap, fernandez2021framework}.

To overcome these challenges, we propose ELASTIC (\textbf{E}vent-\textbf{L}ocation-\textbf{A}gno\textbf{STIC} synchronizer), a trajectory-based synchronization framework that detects both the start and end timestamps of events without relying on annotated event locations. To address the first limitation, ELASTIC infers event timings solely from player and ball trajectories, using features such as player-ball distance and ball acceleration to extract candidate frames that mark potential ball touches. To address the second, we explicitly recover ball receptions by enriching the event sequence with virtual termination events inserted between consecutive events, so that the end of each event is detected jointly with its start. We then find the optimal, order-preserving alignment between the termination-inserted event sequence and the candidate-frame sequence within each in-play segment using an extended Needleman-Wunsch algorithm~\cite{needleman1970general}.

Beyond the methodology, we make two practical contributions. First, we construct a reproducible benchmark by annotating true event timestamps on the Sportec Open DFL Dataset~\cite{bassek2025integrated}, enabling other researchers to reproduce our experiments and use the benchmark in other analyses. This is particularly important because prior studies have relied on proprietary data or indirect evaluation protocols. Second, we go beyond synchronization accuracy itself and evaluate how it affects downstream tasks including next action prediction and pass success prediction~\cite{kim2026better, rahimian2023pass, stockl2021making}. By doing so, we show that synchronization is not merely a technical alignment task but a practically consequential component that materially affects the quality of diverse soccer analytics.

In summary, this paper makes the following contributions:
\begin{itemize}
    \item We propose ELASTIC, a framework for synchronizing event and tracking data in soccer that does not rely on human-annotated event locations, combining candidate frame extraction with an adapted sequence alignment algorithm.
    \item We enable the framework to recover ball receptions by detecting the end of each event as well as its start, through virtual termination events inserted into the event sequence.
    \item We construct a reproducible evaluation benchmark by annotating ground-truth timestamps on a publicly available dataset and show that ELASTIC achieves substantial improvements over existing baselines.
    \item We further demonstrate through downstream task evaluation that improved synchronization leads to practically meaningful gains in soccer analytics.
\end{itemize}

\section{Proposed Framework}
\label{sec:framework}
Given a pair of event and tracking data from a soccer match, our objective is to infer the true start and end timestamps of each on-the-ball event. Formally, the event data is a sequence $E=(e_1, \ldots, e_M)$, where each event specifies its type (e.g., pass, shot, or tackle) and the involved player. The tracking data is a sequence of snapshots $X = (\textbf{x}_1, \ldots, \textbf{x}_T)$ recorded at 10 or 25 frames per second (FPS), where each snapshot contains the 2D position of all players, the 3D position of the ball, and a binary indicator of whether the frame is in play. Given $E$ and $X$, our goal is to find a pair of tracking frames (i.e., snapshot indices) $(t_i^{\text{start}},t_i^{\text{end}})$ corresponding to the true moments when each event $e_i$ was initiated and terminated, while preserving the event order as follows:
\begin{equation}
    1 \le t_1^{\text{start}} \le \cdots \le t_i^{\text{start}} \le t_i^{\text{end}} \le t_{i+1}^{\text{start}} \le \cdots \le t_{M}^{\text{end}} \le T.
\end{equation}

ELASTIC finds these timestamps through five stages: data preprocessing (Section~\ref{sec:preprocess}), selecting candidate frames representing possible ball touches (Section~\ref{sec:candidate}), calculating a compatibility score between every event and candidate frame in each episode (Section~\ref{sec:score}), aligning the event and the candidate-frame sequences using a ``repeat-augmented'' Needleman-Wunsch algorithm (Section~\ref{sec:align}), and postprocessing to refine the alignment (Section~\ref{sec:postprocess}).

\subsection{Data Preprocessing}
\label{sec:preprocess}
First, following Oonk et al.~\cite{oonk2025right}, we split each match into segments of consecutive in-play frames, which we call \emph{episodes}~\cite{kim2023ball}, and align each episode independently. While aligning $E=(e_1, \ldots, e_M)$ and $X = (\textbf{x}_1, \ldots, \textbf{x}_T)$ at once requires the quadratic time complexity of $\mathcal{O}(MT)$, this episode batching substantially reduces this cost to $\mathcal{O}(KM'T')$ where $K \approx 100$ is the number of episodes in a match and $M' \ll M$ and $T' \ll T$ are the number of events and tracking frames in the longest episode, respectively.

Following prior work~\cite{vanroy2023etsy}, we then convert the raw event data into the Soccer Player Action Description Language (SPADL) format~\cite{decroos2019actions} and group events into four categories that share behavior patterns in tracking signals:
\begin{itemize}
    \item \textbf{Open-play outgoing:} \texttt{pass}, \texttt{cross}, \texttt{clearance}, \texttt{shot}, \\ \texttt{shot\_block}, \texttt{keeper\_punch}, \texttt{bad\_touch}
    \item \textbf{Set-piece outgoing:} \texttt{throw\_in}, \texttt{goal\_kick}, \texttt{corner\_short}, \texttt{corner\_crossed}, \texttt{freekick\_short}, \texttt{freekick\_crossed}, \\ \texttt{shot\_freekick}, \texttt{shot\_penalty}
    \item \textbf{Incoming:} \texttt{interception}, \texttt{ball\_recovery}, \texttt{keeper\_save}, \texttt{keeper\_claim}, \texttt{keeper\_pickup}
    \item \textbf{Minor:} \texttt{tackle}, \texttt{dispossessed}
\end{itemize}
A different scoring rule is applied to each category in Section~\ref{sec:score}.

Lastly, to determine both the start and end timestamps of every event through a single alignment per episode, we explicitly insert virtual termination events between consecutive events. Specifically, we conditionally insert one of three virtual events between each pair of adjacent events $(e_i, e_{i+1})$ as follows:
\begin{itemize}
    \item When the current event $e_i$ is a successful shot followed by a kick-off pass, we insert a \texttt{goal} event.
    \item When the next event $e_{i+1}$ is either a throw-in, a goal kick, or a corner kick, we insert an \texttt{out} event.
    \item When $e_i$ and $e_{i+1}$ belong to the same episode and are executed by different players, we insert a \texttt{control} event indicating a ball reception by the next acting player.
    \item Otherwise, we do not insert a virtual event.
\end{itemize}
The resulting enriched sequence alternates as start (original) $\to$ end (inserted) $\to$ start (original) $\to$ end (inserted) $\to \cdots$, allowing every original event's initiation and termination to be determined jointly by a single alignment per episode.

\subsection{Candidate Frame Selection}
\label{sec:candidate}
Oonk et al.~\cite{oonk2025right} first applied the Needleman-Wunsch algorithm~\cite{needleman1970general} to align event and tracking sequences while treating every in-play frame as a potential match for each event. However, this exhaustive treatment has two drawbacks.
First, as tracking data collected from a match at 25 FPS contains about 90,000 in-play frames (i.e., about 60 minutes), filling the dynamic programming (DP) table over all in-play frames is computationally expensive even with dead-ball batching. 
Second, most in-play frames are not physically plausible candidates for on-the-ball events, since the ball is often not sufficiently close to any player or does not exhibit a meaningful change in motion at those moments. Including such frames as matching candidates therefore increases the likelihood of spurious alignments in situations where no event could realistically occur.

To avoid these issues, we extract a sparse subset of \emph{candidate frames} from each episode before the alignment in Section~\ref{sec:align}. Specifically, we consider the following frames as candidates:
\begin{enumerate}
    \item[(a)] local minima of the distance between the ball and a player, capturing the player's potential ball touches;
    \item[(b)] local minima of the distance between the ball and a pitch boundary (i.e., a side line or an end line), capturing the moment the ball goes out of play for \texttt{out} and \texttt{goal} events;
    \item[(c)] local maxima of the ball acceleration, capturing a sudden change in the ball's direction that may not coincide with a distance valley.
\end{enumerate}
Conditions (a) and (b) are evaluated independently for each player and each pitch boundary, and condition (c) is paired with the player closest to the ball at the acceleration peak. We discard any (frame, player) pair whose player-ball distance exceeds \SI{3}{m} or whose ball height exceeds \SI{4}{m}, retaining only those at which a ball touch is physically feasible. Section~\ref{sec:coverage} empirically validates these choices, showing how candidate coverage and final accuracy change as we vary the two thresholds and ablate each detection condition.

\begin{figure}[tb]
\centering
\includegraphics[width=0.45\textwidth]{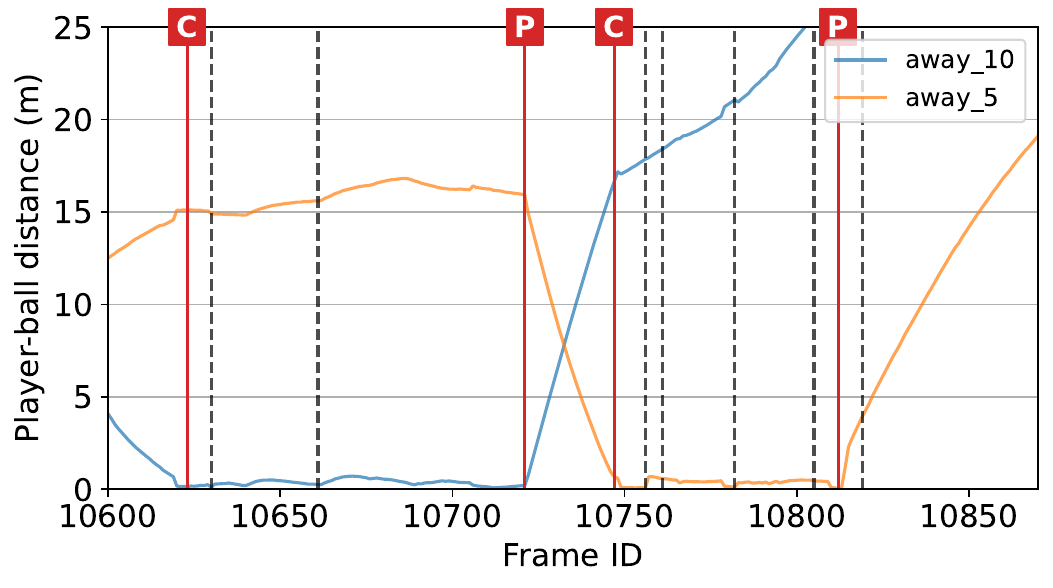}
\vspace{-0.5em}
\caption{Player-ball distances for the two players in a sample interval. Black dashed vertical lines denote candidate frames, and red solid vertical lines indicate those matched to events by the NW alignment, with markers at the top indicating whether each event is a \texttt{control} (C) or a \texttt{pass} (P).}
\label{fig:cand_frames}
\end{figure}

Finally, we group the remaining pairs by their frame index and represent each candidate as $c = (t_c, \mathcal{P}_c)$, where $t_c$ is the frame and $\mathcal{P}_c$ is the set of players and pitch lines associated with $t_c$. Without this grouping, the order-preserving NW alignment in Section~\ref{sec:align} would impose an arbitrary order on these simultaneous candidates, failing to match them with the corresponding co-occurring events whenever the two orderings conflict. See Fig.~\ref{fig:cand_frames} that instantiates the player-ball distances marked with detected candidate frames.

\subsection{Event-Candidate Pairwise Scoring}
\label{sec:score}
In this section, we assign a score in the range from 0 to 1 for every pair of event and candidate in each episode, where a higher score indicates that the event aligns better with the candidate. Without using the human-annotated event locations, we calculate a set of features informative for identifying on-the-ball events only from player and ball trajectories. We compute the score as a weighted sum of per-feature scores, where the features and their weights depend on the event category defined in Section~\ref{sec:preprocess}. Calculating the score of every event-candidate pair yields a pairwise score matrix, which serves as the input to the alignment algorithm in Section~\ref{sec:align}.

Specifically, we first define several per-feature scoring functions for a given candidate frame $c=(t_c,\mathcal{P}_c)$. Each function takes the frame $t_c$ and optionally a target player (or a pitch line) $p \in \mathcal{P}_c$ as input, and returns a value between 0 and 1 based on a clipped linear mapping given by
\begin{equation}
    f(x;x_0,x_1) =
    \begin{cases}
        0 & \text{if } x \le x_0, \\
        1 & \text{if } x \ge x_1, \\
        \frac{x - x_0}{x_1 - x_0} & \text{if } x_0 < x < x_1.
    \end{cases}
\end{equation}
The detailed definition of the scoring functions is as follows:

\vspace{0.5em}
\noindent \textbf{Ball acceleration (BA) score.} Since the ball usually changes its direction when an event occurs, a larger BA should yield a higher score. Thus, for BA $a(t_c)$ at time $t_c$, we define an increasing scoring function as
\begin{equation}
    s_{\text{BA}}(t_c) = f(a(t_c);0,30\,\text{m/s}^2).
    \label{eq:ball_accel}
\end{equation}

\vspace{0.5em}
\noindent \textbf{Player-ball distance (PBD) score.} Since the acting player must be in contact with the ball at an event, a smaller PBD should yield a higher score. Thus, for the distance $d(t,p)$ between player $p$ and the ball at time $t$, we define a decreasing scoring function as
\begin{equation}
    s_{\text{PBD}}(t_c,p) = 1 - f(d(t_c,p);0,3\,\text{m}).
    \label{eq:player_dist}
\end{equation}

\vspace{0.5em}
\noindent \textbf{Kick distance (KD) score.} Since incoming or outgoing events involve the ball traveling a meaningful distance to or from the player, we score candidate $c$ by the maximum PBD over an interval before or after $t_c$ depending on the event category. Specifically, we define the \emph{pre-kick distance} (Pre-KD) by the maximum PBD $\max_{t_c^- \le t \le t_c} d(t, p_c)$ in the interval between $t_c$ and the previous candidate $c^- = (t_c^-, \mathcal{P}_c^-)$ where $p_c \in \mathcal{P}_c^-$, and the \emph{post-kick distance} (Post-KD) by the maximum PBD $\max_{t_c \le t \le t_c^+} d(t, p_c)$ in the interval between $c$ and the next candidate $c^+ = (t_c^+, \mathcal{P}_c^+)$ where $p_c \in \mathcal{P}_c^+$. To distinguish actual ball receptions from subsequent minor touches, we assign a high score to a candidate with a large Pre-KD paired with an incoming event by:
\begin{equation}
    s_{\text{KD}}^-(t_c,p) = f \left( \max_{t_c^- \le t \le t_c} d(t, p); 0, 3\,\text{m} \right).
\end{equation}
Symmetrically, to discriminate kicks against minor touches, we assign a high score to a candidate with a larger Post-KD paired with an outgoing event by:
\begin{equation}
    s_{\text{KD}}^+(t_c,p) = f \left( \max_{t_c \le t \le t_c^+} d(t, p); 0, 3\,\text{m} \right).
\end{equation}
We include either $s_{\text{KD}}^-$ or $s_{\text{KD}}^+$ in the total score defined in Eq.~\ref{eq:outgoing}--\ref{eq:minor}, depending on the type of the paired event.

\vspace{0.5em}
\noindent \textbf{Player-ball distance slope (PBDS) score.} Since the PBD typically exhibits a valley at the moment of an event, we penalize candidates that deviate from this V-shaped pattern using the \emph{pre-slope} $v^-(t,p)$ and \emph{post-slope} $v^+(t,p)$ of the PBD, defined as:
\begin{equation}
    \small
    v^-(t, p) = \frac{d(t, p) - d(t - h, p)}{h}, \quad
    v^+(t, p) = \frac{d(t + h, p) - d(t, p)}{h} \nonumber
\end{equation}
where $h=0.2\,\text{s}$. A positive pre-slope indicates that the ball is already moving away from $p$ before $t_c$, which is incompatible with an outgoing event. We therefore penalize candidate $c$ paired with an outgoing event by:
\begin{equation}
    s_{\text{PBDS}}^-(t_c,p) = 1 - f(v^-(t_c, p); 0, 7\,\text{m/s}).
\end{equation}
Likewise, a negative post-slope implies that the ball is still approaching $p$ after $t_c$, which is incompatible with an incoming event. We thus penalize $c$ paired with an incoming event by:
\begin{equation}
    s_{\text{PBDS}}^+(t_c,p) = f(v^+(t_c, p); -7\,\text{m/s}, 0).
\end{equation}
We include either $s_{\text{PBDS}}^-$, $s_{\text{PBDS}}^+$, or $s_{\text{OD}}$ defined below into the total score as in Eq.~\ref{eq:outgoing}--\ref{eq:minor}, depending on the type of the paired event.

\vspace{0.5em}
\noindent \textbf{Opponent distance (OD) score.} Since two players contest the ball when a \texttt{tackle} or a \texttt{dispossessed} event occurs, the ball should be also close to the nearest opposing player as well as the acting player. We thus assign a high score to such candidates by defining the scoring function as:
\begin{equation}
    s_{\text{OD}}(t_c,p) = 1 - f \left( \min_{q \in O(p)} d(t_c, q); 0, 3\,\text{m} \right),
    \label{eq:oppo_dist}
\end{equation}
where $O(p)$ is the set of players in the opposing team against $p$.

\begin{figure}[tb]
    \centering
    \begin{subfigure}[t]{\columnwidth}
        \centering
        \includegraphics[width=0.95\columnwidth]{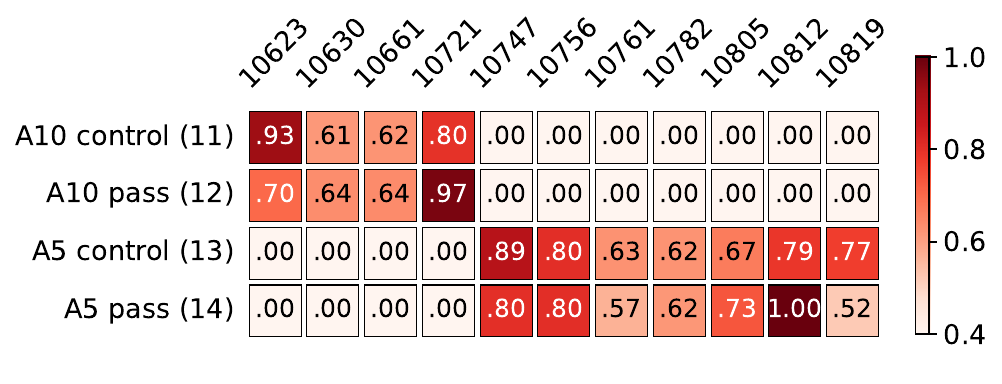}
        \vspace{-1em}
        \caption{Pairwise scores $s(e_i, c_j)$}
        \label{fig:score_mat}
    \end{subfigure}
    \vspace{1em}
    \begin{subfigure}[t]{\columnwidth}
        \centering
        \includegraphics[width=0.98\columnwidth]{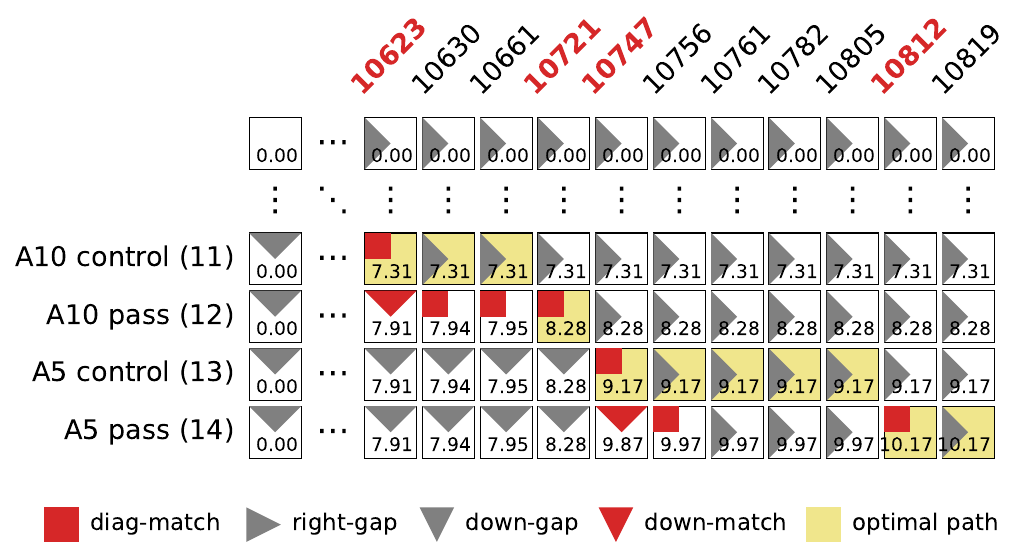}
        \vspace{-0.3em}
        \caption{DP table $F_{i,j}$ with backpointer moves $B_{i,j}$}
        \label{fig:dp_table}
    \end{subfigure}
    \vspace{-1.5em}
    \caption{Components of the NW alignment for the four events of the interval shown in Fig.~\ref{fig:cand_frames}. The match moves (diag-match or down-match) along the yellow-highlighted optimal path in (b) correspond to the final event-frame pairs.}
    \label{fig:nw}
\end{figure}

\vspace{0.5em}
Then, we compute the score $s(e, c)$ between every pair of an event $e$ executed by player $p_e$ and a candidate $c = (t_c, \mathcal{P}_c)$ per episode as a weighted sum of feature scores defined above. We require $p_e \in \mathcal{P}_c$ as a hard constraint, setting $s(e, c) = 0$ regardless of the other features if $p_e \notin \mathcal{P}_c$. That is, the total score defined in Eq.~\ref{eq:outgoing}--\ref{eq:minor} is evaluated only when $p_e \in \mathcal{P}_c$.

We include different component scores depending on the category of $e$ (defined in Section~\ref{sec:preprocess}). If $e$ is an outgoing event in either an open play or a set piece, the ball should depart from the executing player and travel a meaningful distance after the event, so we use the post-KD score $s_{\text{KD}}^+$ and the pre-slope score $s_{\text{PBDS}}^-$:
\begin{multline}
    s(e, c) = \lambda_{\text{BA}} s_{\text{BA}}(t_c) + \lambda_{\text{PBD}} s_{\text{PBD}}(t_c, p_e) \\
    + \lambda_{\text{KD}} s_{\text{KD}}^+(t_c, p_e) + \lambda_{\text{PBDS}} s_{\text{PBDS}}^-(t_c, p_e).
\label{eq:outgoing}
\end{multline}
In contrast, if $e$ is an incoming event, the ball should approach $p_e$ from a distant location before the event, so we use the pre-KD and post-slope counterparts $s_{\text{KD}}^-$ and $s_{\text{PBDS}}^+$:
\begin{multline}
    s(e, c) = \lambda_{\text{BA}} s_{\text{BA}}(t_c) + \lambda_{\text{PBD}} s_{\text{PBD}}(t_c, p_e) \\
    + \lambda_{\text{KD}} s_{\text{KD}}^-(t_c, p_e) + \lambda_{\text{PBDS}} s_{\text{PBDS}}^+(t_c, p_e).
\label{eq:incoming}
\end{multline}
When a minor event occurs, two players contest the ball at close range, so we use the OD score instead of the PBDS scores. For the KD term, \texttt{tackle} uses $s_{\text{KD}}^-$ since the ball travels toward the tackling player before they touch it, while \texttt{dispossessed} uses $s_{\text{KD}}^+$ since the executing player loses the ball after the event. The score is thus defined as:
\begin{multline}
    s(e, c) = \lambda_{\text{BA}} s_{\text{BA}}(t_c) + \lambda_{\text{PBD}} s_{\text{PBD}}(t_c, p_e) \\
    + \lambda_{\text{KD}} s_{\text{KD}}(t_c, p_e) + \lambda_{\text{OD}} s_{\text{OD}}(t_c, p_e),
\label{eq:minor}
\end{multline}
where $s_{\text{KD}} = s_{\text{KD}}^-$ for \texttt{tackle} and $s_{\text{KD}} = s_{\text{KD}}^+$ for \texttt{dispossessed}. We set all the weights to $\lambda_{\text{BA}} = \lambda_{\text{PBD}} = \lambda_{\text{KD}} =  \lambda_{\text{PBDS}} = \lambda_{\text{OD}} = 0.25$, so that each $s(e, c)$ falls between 0 and 1.

In Section~\ref{sec:sensitivity}, we examine the sensitivity of the final accuracy to these hyperparameters by varying each clipping threshold and weight of the per-feature scoring functions, and conduct an ablation study that confirms the necessity of each feature score by removing one term at a time. Fig.~\ref{fig:score_mat} illustrates the resulting score matrix for sample events, where each row corresponds to an event and each column corresponds to a candidate frame.

\subsection{Repeat-Augmented Needleman-Wunsch Algorithm}
\label{sec:align}
Using the pairwise score matrix defined in Section~\ref{sec:score}, we align the event sequence and the candidate frame sequence within each episode while preserving their order. To this end, we adapt the Needleman-Wunsch (NW) algorithm~\cite{needleman1970general}, which was originally proposed in bioinformatics to find the globally optimal alignment between two amino-acid or DNA sequences. As the two sequences usually differ in length and cannot be matched one-to-one, NW introduces a \emph{gap penalty} that controls which elements to leave as gaps on either side instead of forcing them into a match. In our framework, gaps on both sides naturally absorb ambiguous events that have no compatible candidate frame, as well as candidate frames at which no event actually occurs.

\vspace{0.5em} \noindent \textbf{Vanilla NW.}
For an episode $k$, let $E_k = (e_{k,1}, \ldots, e_{k,m_k})$ and $C_k = (c_{k,1}, \ldots, c_{k,n_k})$ denote the enriched event sequence and the candidate frame sequence, respectively. For notational simplicity, we drop the episode index $k$ and write $E_k = (e_1, \ldots, e_m)$ and $C_k = (c_1, \ldots, c_n)$ throughout this section. The vanilla NW initializes a dynamic programming (DP) table $F \in \mathbb{R}^{(m+1) \times (n+1)}$ as $F_{0,0} = 0$, $F_{i,0} = i \cdot g_e$, and $F_{0,j} = j \cdot g_c$, where $g_e$ and $g_c$ are the gap penalties for leaving an event unmatched and a candidate frame unused, respectively. The remaining entries are then filled recursively by taking the maximum over three possible moves at each cell:
\begin{equation}
    F_{i,j} = \max \begin{cases}
    F_{i-1, j-1} + s(e_i, c_j), & \text{(diag-match)} \\
    F_{i, j-1} + g_c, & \text{(right-gap: $c_j$ unmatched)} \\
    F_{i-1, j} + g_e, & \text{(down-gap: $e_i$ unmatched)}
    \end{cases}
    \label{eq:nw}
\end{equation}
where $s(e_i, c_j)$ is the pairwise score defined in Section~\ref{sec:score}, while recording the selected move in a backpointer table
\begin{equation}
    B \in\{\text{diag-match}, \text{right-gap}, \text{down-gap}\}^{(m+1) \times (n+1)}. \nonumber
\end{equation}
Once $F$ is completed, we trace back from the last position $(m, n)$ to the origin $(0, 0)$ along the opposite directions stored in $B$, and the resulting path yields the optimal alignment between $E_k$ and $C_k$.

\vspace{0.5em} \noindent \textbf{Extension with a down-match move.}
The threefold recurrence of this vanilla NW enforces a one-to-one correspondence in which each candidate frame can match at most one event. In soccer, however, two consecutive events frequently correspond to the same moment. For example, a loss of possession by one player and the gain of possession by an opposing player are separately recorded as \texttt{dispossessed} and \texttt{tackle}, but actually describe a single ball contest. Another typical example is a one-touch action such as a \texttt{control} followed by a \texttt{pass} or \texttt{shot}, in which the player makes a pass or a shot at their first contact with the ball. To accommodate such cases, we extend the NW recurrence in Eq.~\ref{eq:nw} with a \emph{down-match} option, which goes downward like a down-gap but matches the current event again to the previously matched candidate frame instead of leaving it as a gap:
\begin{equation}
    F_{i,j} = \max \begin{cases}
    F_{i-1, j-1} + s(e_i, c_j), & \text{(diag-match)} \\
    F_{i, j-1} + g_c, & \text{(right-gap)} \\
    F_{i-1, j} + g_e, & \text{(down-gap)} \\
    F_{i-1, j} + s(e_i, c_j) + r, & \text{(down-match)}
    \end{cases}
    \label{eq:ranw}
\end{equation}
where $r \le 0$ is a constant \emph{repeat penalty}. A down-match move duplicates the match of $c_j$, assigning it to both $e_{i-1}$ and $e_i$ at the cost of $r$, which indicates that the two events happen at the same moment. Like the vanilla NW algorithm, we record in the backpointer table
\begin{equation}
    B \in \{\text{diag-match}, \text{right-gap}, \text{down-gap}, \text{down-match}\}^{(m+1) \times (n+1)} \nonumber
\end{equation}
indicating which of the four moves attained the maximum.

\vspace{0.5em} \noindent \textbf{Penalty design.}
The gap and repeat penalties are designed based on domain-specific intuitions. First, we set the candidate frame gap penalty to $g_c = 0$ so that the alignment is driven solely by the pairwise scores. This is because the number of candidate frames $n$ is generally larger than the number of events $m$, and at least $n - m$ frames must remain unmatched in any alignment. By setting $g_c = 0$, the algorithm determines which candidates to leave out only based on the pairwise score $s(\cdot)$. Likewise, we set the event gap penalty to $g_e = 0$ to allow an event to remain unmatched when no compatible candidate exists, deferring the rejection of low-confidence matches to the postprocessing threshold in Section~\ref{sec:postprocess}. Lastly, we set the repeat penalty to $r = -0.1$, so that two consecutive events are assigned to the same frame only when their scores are high enough to outweigh the penalty. Section~\ref{sec:sensitivity} shows that the alignment is insensitive to $g_e$ and validates the choices of $g_c$ and $r$.

\vspace{0.5em} \noindent \textbf{Traceback.}
Once $F$ is fully populated, we obtain the optimal alignment by tracing back along $B$ from $(m, n)$ to $(0, 0)$. Each move recorded in $B_{i,j}$ shifts the indices in a different way: a diag-match shifts $(i, j)$ to $(i-1, j-1)$ while matching $e_i$ to $c_j$; a right-gap shifts to $(i, j-1)$ while leaving $c_j$ unmatched; and a down-gap shifts to $(i-1, j)$ while leaving $e_i$ unmatched. A down-match shifts in the same direction as a down-gap but matches $e_i$ to $c_j$ rather than leaving $e_i$ as a gap. When the trace reaches $(0, 0)$, every entry of the enriched event sequence is assigned either to a candidate frame or to a gap. Fig.~\ref{fig:dp_table} shows the resulting DP table and the optimal alignment for the sample events used in Fig.~\ref{fig:score_mat}.

\subsection{Postprocessing}
\label{sec:postprocess}
After obtaining the optimal alignment between the enriched event sequence and the candidate frame sequence, we apply three postprocessing steps to derive the final output. First, when an adjacent dispossessed-tackle pair is matched to different candidate frames, we unify them by snapping both events to the frame with the higher pairwise score, since the two events describe a single ball contest. Second, we reject any match whose pairwise score falls below 0.5 by marking it as unsynchronized, leaving low-confidence matches empty rather than contaminating downstream analyses. Finally, we fold the alignment over the enriched event sequence back into the original events, where each original event $e_i$ takes its start timestamp from its matched frame, and its end timestamp from the candidate frame matched to the virtual termination event inserted immediately after $e_i$ in Section~\ref{sec:preprocess}.

\section{Main Experiments}
\label{sec:experiments}
For rigorous evaluation, we first re-annotated the event timestamps in a public dataset to construct a reliable ground-truth benchmark, and validated its quality via a cross-annotator comparison (Section~\ref{sec:benchmark}). We then compared the synchronization accuracy of ELASTIC on this benchmark (Section~\ref{sec:accuracy}) against several baseline methods described in Section~\ref{sec:baseline}. To further examine our methodological design, we measured the coverage of the extracted candidate frames over the true event timestamps (Section~\ref{sec:coverage}), conducted a sensitivity study on the hyperparameters (Section~\ref{sec:sensitivity}), and analyzed the runtime of the methods (Section~\ref{sec:runtime}).

\subsection{Ground-Truth Benchmark Construction}
\label{sec:benchmark}
To construct a reproducible benchmark, we re-annotated the event timestamps in the Sportec Open DFL Dataset~\cite{bassek2025integrated}, a publicly available collection of event and tracking data from seven matches of the German Bundesliga's first and second divisions. For effective annotation, we developed a JavaScript-based annotation tool that replays the match animation alongside the event records, allowing an annotator to inspect each event and correct its timestamp\footnote{The tool with the annotation results is available at \url{https://elastic-annotator.com}. The annotator can jump the video to its recorded moment by clicking an event, and can reassign the event's timestamp to the currently displayed frame with a single click.}.

Using this tool, three authors with expertise in soccer analytics independently re-labeled the event timestamps from three of the matches at 25 FPS (\SI{0.04}{s} resolution) to align with the tracking data. Since the original dataset does not record event terminations, we further inserted \texttt{control}, \texttt{out}, and \texttt{goal} events with their true timestamps wherever they were missing. 

\begin{table}[t]
\centering
\caption{Agreement statistics of event time labels across three annotators. ``MD'' denotes the mean pairwise absolute difference in frames between the labeled timestamps for each event. ``Exact3'' and ``Exact2'' indicate the number of events where at least three or two annotators, respectively, provided exactly the same timestamp. ``Close3'' and ``Close2'' indicate the number of events where at least three or two labels, respectively, fall within a two-frame (\SI{0.08}{s}) window.}
\label{tab:annot}
\setlength{\tabcolsep}{1.5pt}
\footnotesize
\begin{tabular}{l r r r r r r}
\toprule
\textbf{Category} & \multicolumn{1}{c}{\textbf{Total}} & \multicolumn{1}{c}{\textbf{MD}} & \multicolumn{1}{c}{\textbf{Exact3}} & \multicolumn{1}{c}{\textbf{Exact2}} & \multicolumn{1}{c}{\textbf{Close3}} & \multicolumn{1}{c}{\textbf{Close2}} \\
\midrule
Outgoing-OP & 2,690 & 0.279 & 2,615 (97.2\%) & 2,684 (99.8\%) & 2,647 (98.4\%) & 2,688 (99.9\%) \\
Outgoing-SP  & 282   & 0.021 & 279 (98.9\%)   & 282 (100.0\%)  & 280 (99.3\%)   & 282 (100.0\%) \\
Incoming      & 295   & 1.530 & 266 (90.2\%)   & 288 (97.6\%)   & 274 (92.9\%)   & 290 (98.3\%) \\
Minor         & 231   & 3.873 & 189 (81.8\%)   & 223 (96.5\%)   & 190 (82.3\%)   & 227 (98.3\%) \\
\midrule
Event start & 3,498 & 0.601 & 3,349 (95.7\%) & 3,477 (99.4\%) & 3,391 (96.9\%) & 3,487 (99.7\%) \\
Event end   & 2,972 & 0.426 & 2,799 (94.2\%) & 2,911 (97.9\%) & 2,859 (96.2\%) & 2,938 (98.9\%) \\
\midrule
Total       & 6,470 & 0.521 & 6,148 (95.0\%) & 6,388 (98.7\%) & 6,250 (96.6\%) & 6,425 (99.3\%) \\
\bottomrule
\end{tabular}
\vspace{-0.3em}
\end{table}

Table~\ref{tab:annot} reports the agreement statistics across the three annotators. All annotators assigned exactly the same timestamp for 95.0\% of events, with a mean pairwise difference (MD) of 0.521 frames (\SI{0.0208}{s}). This high consistency stems from the meticulous re-annotation process, where each annotator paused the animation at every event to pinpoint its exact frame. Moreover, for 99.3\% of events, at least two annotations fall within two frames (\SI{0.08}{s}) of each other. Assuming that such close agreement reflects reliability, this result implies that the median of the three timestamps provides a reliable estimate for 99.3\% of events. Based on this observation, we adopt the median of the annotated timestamps for each event as the ground truth in the subsequent experiments.

\subsection{Baselines and Evaluation Metrics}
\label{sec:baseline}
We compare ELASTIC against three baselines. ETSY~\cite{vanroy2023etsy} is a rule-based synchronizer that processes events in chronological order, assigning each event the frame that minimizes the sum of pairwise distances among the acting player, the ball, and the annotated event location. Biermann et al.~\cite{biermann2023synchronization} proposed a learning-based synchronizer, which slides a window over the tracking data and classifies whether each window's center frame corresponds to an event based on aggregated window features. Since their method requires training data with ground-truth event timestamps, we evaluate it via leave-one-match-out cross-validation, training on two of the three benchmark matches and testing on the remaining match. As it detects only pass-like (outgoing) events, we leave incoming and minor events unsynchronized. DataBallPy~\cite{oonk2025right, oonk2026databallpy} finds a globally optimal alignment using the NW algorithm as our framework, but aligns events to all in-play frames without candidate frame selection. It synchronizes only \texttt{pass}, \texttt{shot}, and \texttt{tackle} events, so we map each outgoing event to a \texttt{shot} if its type is \texttt{shot}, \texttt{shot\_freekick}, or \texttt{shot\_penalty} (among all types listed in Section~\ref{sec:preprocess}), and to a \texttt{pass} otherwise. Since DataBallPy does not handle ball receptions, it cannot synchronize incoming events. Among minor events, it covers only \texttt{tackle}, leaving \texttt{dispossessed} unsynchronized.

\begin{table*}[t]
\centering
\caption{Synchronization accuracy of three baselines and two ELASTIC variants across event categories.}
\vspace{-0.5em}
\label{tab:accuracy}
\setlength{\tabcolsep}{3pt}
\footnotesize
\begin{tabular}{l l | r r | @{\hspace{6pt}} r r r r r}
\toprule
\textbf{Category} & \textbf{Method} & \multicolumn{1}{c}{\textbf{Total}} & \multicolumn{1}{c| @{\hspace{6pt}}}{\textbf{MD}} & \multicolumn{1}{c}{\textbf{W2}} & \multicolumn{1}{c}{\textbf{W5}} & \multicolumn{1}{c}{\textbf{W25}} & \multicolumn{1}{c}{\textbf{W50}} & \multicolumn{1}{c}{\textbf{Valid}} \\
\specialrule{1pt}{2pt}{3pt}
Open-play & ETSY & 2,690 & 10.076 & 1,510 (56.1\%) & 1,853 (68.9\%) & 2,219 (82.5\%) & 2,365 (87.9\%) & 2,484 (92.3\%) \\
outgoing & Biermann et al. &  & 15.268 & 2,103 (78.2\%) & 2,184 (81.2\%) & 2,278 (84.7\%) & 2,397 (89.1\%) & \textbf{2,690 (100.0\%)} \\
 & DataBallPy &  & 4.381 & 2,313 (86.0\%) & 2,533 (94.2\%) & 2,593 (96.4\%) & 2,641 (98.2\%) & 2,687 (99.9\%) \\
 & ELASTIC-Greedy &  & 7.251 & 2,315 (86.1\%) & 2,322 (86.3\%) & 2,349 (87.3\%) & 2,380 (88.5\%) & 2,525 (93.9\%) \\
 & ELASTIC-NW &  & \textbf{1.050} & \textbf{2,631 (97.8\%)} & \textbf{2,642 (98.2\%)} & \textbf{2,652 (98.6\%)} & \textbf{2,666 (99.1\%)} & 2,686 (99.9\%) \\
\midrule
Set-piece & ETSY & 282 & 28.560 & 183 (64.9\%) & 196 (69.5\%) & 198 (70.2\%) & 201 (71.3\%) & 230 (81.6\%) \\
outgoing & Biermann et al. &  & 3.915 & 244 (86.5\%) & 259 (91.8\%) & 271 (96.1\%) & 274 (97.2\%) & \textbf{282 (100.0\%)} \\
 & DataBallPy &  & 3.280 & 264 (93.6\%) & 272 (96.5\%) & 273 (96.8\%) & 276 (97.9\%) & \textbf{282 (100.0\%)} \\
 & ELASTIC-Greedy &  & 3.594 & 268 (95.0\%) & 272 (96.5\%) & 272 (96.5\%) & 273 (96.8\%) & 280 (99.3\%) \\
 & ELASTIC-NW &  & \textbf{0.060} & \textbf{276 (97.9\%)} & \textbf{277 (98.2\%)} & \textbf{277 (98.2\%)} & \textbf{277 (98.2\%)} & 277 (98.2\%) \\
\midrule
Incoming & ETSY & 295 & 23.287 & 135 (45.8\%) & 165 (55.9\%) & 200 (67.8\%) & 221 (74.9\%) & 258 (87.5\%) \\
 & Biermann et al. &  & --- & 0 (0.0\%) & 0 (0.0\%) & 0 (0.0\%) & 0 (0.0\%) & 0 (0.0\%) \\
 & DataBallPy &  & --- & 0 (0.0\%) & 0 (0.0\%) & 0 (0.0\%) & 0 (0.0\%) & 0 (0.0\%) \\
 & ELASTIC-Greedy &  & 22.312 & 214 (72.5\%) & 216 (73.2\%) & 222 (75.3\%) & 236 (80.0\%) & 270 (91.5\%) \\
 & ELASTIC-NW &  & \textbf{5.799} & \textbf{266 (90.2\%)} & \textbf{268 (90.8\%)} & \textbf{273 (92.5\%)} & \textbf{283 (95.9\%)} & \textbf{294 (99.7\%)} \\
\midrule
Minor & ETSY & 231 & 18.370 & 76 (32.9\%) & 102 (44.2\%) & 165 (71.4\%) & 192 (83.1\%) & 215 (93.1\%) \\
 & Biermann et al. &  & --- & 0 (0.0\%) & 0 (0.0\%) & 0 (0.0\%) & 0 (0.0\%) & 0 (0.0\%) \\
 & DataBallPy &  & 15.284 & 39 (16.9\%) & 59 (25.5\%) & 95 (41.1\%) & 112 (48.5\%) & 120 (51.9\%) \\
 & ELASTIC-Greedy &  & 13.322 & 144 (62.3\%) & 154 (66.7\%) & 180 (77.9\%) & 194 (84.0\%) & 214 (92.6\%) \\
 & ELASTIC-NW &  & \textbf{2.883} & \textbf{203 (87.9\%)} & \textbf{210 (90.9\%)} & \textbf{222 (96.1\%)} & \textbf{226 (97.8\%)} & \textbf{231 (100.0\%)} \\
\specialrule{1pt}{2pt}{3pt}
Event start & ETSY & 3,498 & 13.030 & 1,904 (54.4\%) & 2,316 (66.2\%) & 2,782 (79.5\%) & 2,979 (85.2\%) & 3,187 (91.1\%) \\
 & Biermann et al. &  & 14.201 & 2,347 (67.1\%) & 2,443 (69.8\%) & 2,549 (72.9\%) & 2,671 (76.4\%) & 2,972 (85.0\%) \\
 & DataBallPy &  & 4.703 & 2,616 (74.8\%) & 2,864 (81.9\%) & 2,961 (84.6\%) & 3,029 (86.6\%) & 3,089 (88.3\%) \\
 & ELASTIC-Greedy &  & 8.572 & 2,941 (84.1\%) & 2,964 (84.7\%) & 3,023 (86.4\%) & 3,083 (88.1\%) & 3,289 (94.0\%) \\
 & ELASTIC-NW &  & \textbf{1.495} & \textbf{3,376 (96.5\%)} & \textbf{3,397 (97.1\%)} & \textbf{3,424 (97.9\%)} & \textbf{3,452 (98.7\%)} & \textbf{3,488 (99.7\%)} \\
\midrule
Event end & ELASTIC-Greedy & 2,972 & 13.092 & 2,313 (77.8\%) & 2,334 (78.5\%) & 2,381 (80.1\%) & 2,449 (82.4\%) & 2,622 (88.2\%) \\
 & ELASTIC-NW &  & \textbf{1.937} & \textbf{2,782 (93.6\%)} & \textbf{2,803 (94.3\%)} & \textbf{2,853 (96.0\%)} & \textbf{2,890 (97.2\%)} & \textbf{2,935 (98.8\%)} \\
\specialrule{1pt}{2pt}{3pt}
Total & ELASTIC-Greedy & 6,470 & 10.577 & 5,254 (81.2\%) & 5,298 (81.9\%) & 5,404 (83.5\%) & 5,532 (85.5\%) & 5,911 (91.4\%) \\
 & ELASTIC-NW &  & \textbf{1.697} & \textbf{6,158 (95.2\%)} & \textbf{6,200 (95.8\%)} & \textbf{6,277 (97.0\%)} & \textbf{6,342 (98.0\%)} & \textbf{6,423 (99.3\%)} \\
\bottomrule
\end{tabular}
\end{table*}

In addition to the three baselines, we evaluate two variants of our framework: \emph{ELASTIC-NW}, the full model using the NW alignment, and \emph{ELASTIC-Greedy}, an ablated variant that shares the candidate frame selection (Section~\ref{sec:candidate}) and the pairwise scoring (Section~\ref{sec:score}) with ELASTIC-NW but replaces the NW alignment (Section~\ref{sec:align}) with greedy matching. Specifically, when an episode begins with a set piece, ELASTIC-Greedy first locates the restarting kick by selecting the first candidate frame that is paired with the kicker and whose post-slope exceeds \SI{7}{m/s}. It then processes the remaining events in chronological order, assigning each event the highest-scoring candidate between the frame assigned to the previous event and \SI{5}{s} after the annotated timestamp of the current event. Once all original events are synchronized, each virtual termination event (i.e., \texttt{goal}, \texttt{out}, or \texttt{control}) is detected between the frames assigned to its two neighboring original events, and low-confidence matches are rejected by the same score threshold as in Section~\ref{sec:postprocess}.

We report the mean absolute difference (MD) in frames between the predicted and ground-truth timestamps, the number of events whose predicted frame falls within 2, 5, 25, and 50 frames of the ground truth (W2, W5, W25, W50), and the number of events to which the synchronizer successfully assigns a frame (Valid). Among these, we adopt W2 (a two-frame, \SI{0.08}{s} tolerance) as our primary metric rather than exact alignment, so as not to penalize one- or two-frame discrepancies unrelated to actual synchronization accuracy. Such tiny differences typically arise from differing definitions of event moments across providers or annotators, or from local extrema shifted by one or two frames after the tracking data is smoothed. If only exact alignment were counted as correct, these artifacts would be conflated with genuine synchronization errors. We therefore treat a prediction within two frames of the ground truth as accurate, which remains reliable since the annotators' labels agree within two frames for 99.3\% of events (Section~\ref{sec:benchmark}).

\subsection{Synchronization Accuracy}
\label{sec:accuracy}
As shown in Table~\ref{tab:accuracy}, ELASTIC-NW substantially outperforms the baselines across all event categories. For event starts (i.e., moments of the original events), it achieves 96.5\% W2 accuracy, compared to 54.4\% (ETSY), 67.1\% (Biermann et al.), and 74.8\% (DataBallPy). While this gap partly reflects that Biermann et al. and DataBallPy cover only a subset of event types, ELASTIC-NW remains the most accurate even on outgoing events in both open play (97.8\% vs.\ 86.0\% of DataBallPy) and set pieces (97.9\% vs.\ 93.6\% of DataBallPy), which all baselines are designed to cover. Moreover, it also detects event ends with 93.6\% W2 accuracy, which are unrecorded ball receptions that prior methods leave unsynchronized. As a result, ELASTIC-NW synchronizes 95.2\% of all events within two frames of the ground truth, with a mean difference of only 1.697 frames (\SI{0.068}{s}).

The comparison between the two ELASTIC variants highlights the benefit of global alignment. Although ELASTIC-Greedy uses the same candidate frames, its sequential matching propagates early errors to later events, yielding markedly lower accuracy on event starts (84.1\% vs. 96.5\%). The gap is most pronounced for minor events (62.3\% vs. 87.9\%), whose behavior patterns are more ambiguous than those of other categories. The order-preserving NW alignment resolves such ambiguities jointly, preventing the local mistakes that accumulate under greedy matching. Overall, the precise and robust synchronization achieved by ELASTIC-NW makes it a reliable foundation for downstream tasks where accurate event timing is essential.

\subsection{Candidate Frame Coverage}
\label{sec:coverage}
Since ELASTIC aligns events only to the extracted candidate frames, the candidate selection stage caps the achievable accuracy: if no candidate lies near the true frame of an event, no subsequent alignment can synchronize it correctly. To quantify this cap, we define the \emph{W2 coverage} of a candidate set as the fraction of ground-truth event timestamps for which a candidate frame associated with the acting player exists within two frames. The W2 coverage equals the W2 accuracy of an oracle that always selects the correct candidate, and thus upper-bounds ELASTIC-NW and any other synchronizer operating on the same candidates.

\begin{table}[htb]
\caption{W2 coverage of the extracted candidates over the true event frames and W2 accuracy of ELASTIC-NW under different detection conditions and thresholds, along with the number of candidates and the per-match runtime (mean $\pm$ std). $\star$ marks the default configuration, and cells are shaded by the size of the drop from the default setting: yellow for at least 1 pp, light red for 5 pp, or dark red for 15 pp.}
\label{tab:coverage}
\setlength{\tabcolsep}{4pt}
\footnotesize
\begin{tabular}{@{}l|rcccc|r@{}}
\toprule
& & \multicolumn{3}{c}{\textbf{Event coverage}} & & \\
\cmidrule(lr){3-5}
\textbf{Config.} & \textbf{\#Cand.} & \textbf{Start} & \textbf{End}
& \textbf{Total} & \textbf{Acc.} & \textbf{Time (s)} \\
\midrule
\multicolumn{7}{@{}l}{\textbf{Detection conditions}} \\
\midrule
\quad (a)+(b) only        & 10,926 & \cd 76.9\% & \cd 77.5\% & \cd 77.2\% & \cd 75.4\% & $20.09 \pm 1.38$ \\
\quad (c) only            & 14,363 & \cl 90.7\% & \cl 86.0\% & \cl 88.5\% & \cl 85.4\% & $35.30 \pm 4.54$ \\
\quad (a)+(b)+(c)$^\star$ & 16,979 & 98.5\% & 98.2\% & 98.4\% & 95.2\% & $37.62 \pm 5.25$ \\
\midrule
\multicolumn{7}{@{}l}{\textbf{Player-ball distance threshold}} \\
\midrule
\quad \SI{1}{m}           & 10,352 & \cl 93.1\% & \cy 95.1\% & \cy 94.0\% & \cl 89.4\% & $33.85 \pm 4.42$ \\
\quad \SI{2}{m}           & 13,716 & 98.4\% & 98.2\% & 98.3\% & 94.9\% & $34.60 \pm 5.29$ \\
\quad \SI{3}{m}$^\star$   & 16,979 & 98.5\% & 98.2\% & 98.4\% & 95.2\% & $37.62 \pm 5.25$ \\
\quad \SI{4}{m}           & 20,516 & 98.6\% & 98.2\% & 98.4\% & 95.1\% & $36.50 \pm 5.23$ \\
\quad \SI{5}{m}           & 24,116 & 98.6\% & 98.2\% & 98.4\% & 95.2\% & $37.46 \pm 5.18$ \\
\quad $\infty$            & 220,812 & 98.7\% & 98.2\% & 98.5\% & 95.1\% & $115.82 \pm 24.40$ \\
\midrule
\multicolumn{7}{@{}l}{\textbf{Ball height threshold}} \\
\midrule
\quad \SI{1}{m}           & 14,306 & \cl 83.9\% & \cl 84.5\% & \cl 84.2\% & \cd 77.2\% & $32.80 \pm 4.44$ \\
\quad \SI{2}{m}           & 15,685 & \cl 93.3\% & \cy 93.8\% & \cy 93.5\% & \cl 88.5\% & $32.30 \pm 5.55$ \\
\quad \SI{3}{m}           & 16,639 & 98.5\% & 98.2\% & 98.4\% & 95.1\% & $33.90 \pm 4.04$ \\
\quad \SI{4}{m}$^\star$   & 16,979 & 98.5\% & 98.2\% & 98.4\% & 95.2\% & $37.62 \pm 5.25$ \\
\quad \SI{5}{m}           & 17,202 & 98.5\% & 98.2\% & 98.4\% & 95.2\% & $32.61 \pm 4.84$ \\
\quad $\infty$            & 17,726 & 98.5\% & 98.3\% & 98.4\% & 95.2\% & $32.77 \pm 4.44$ \\
\bottomrule
\end{tabular}
\end{table}

Table~\ref{tab:coverage} reports the W2 coverage and the W2 synchronization accuracy of ELASTIC-NW under different candidate selection settings. With the default configuration, the extracted candidates cover 98.4\% of the ground-truth event timestamps, implying that nearly every event has a correct candidate to align with. The first block of the table shows that the detection conditions (a)--(c) defined in Section~\ref{sec:candidate} are complementary: using only the distance conditions (a) and (b) or only the acceleration condition (c) drops the coverage to 77.2\% and 88.5\%, respectively. The former misses touches at which the player-ball distance forms no local minimum, while the latter misses soft touches without a sudden change in ball motion.

The remaining blocks vary the two feasibility thresholds. Tight thresholds miss legitimate touches: a \SI{1}{m} distance limit discards true touches whose measured player-ball distance is inflated by trajectory noise, and a height limit below \SI{3}{m} loses aerial touches, both collapsing the coverage to below 95\%. Once both thresholds reach \SI{3}{m}, the coverage and accuracy saturate, while removing the limits substantially increases the number of extracted candidates and thus the computation time. We therefore set the distance threshold to \SI{3}{m} and the height threshold to \SI{4}{m}, where the extraction stays efficient without performance drops and the retained candidates remain physically plausible for an actual ball touch.

\subsection{Hyperparameter Sensitivity}
\label{sec:sensitivity}
ELASTIC involves three groups of hyperparameters: the feature weights in Eq.~\ref{eq:outgoing}--\ref{eq:minor}, the clipping bounds of the per-feature scoring functions in Eq.~\ref{eq:ball_accel}--\ref{eq:oppo_dist}, and the alignment penalties in Eq.~\ref{eq:ranw}. To justify our choice of the hyperparameters, we vary each of them while fixing the others to their defaults and measure the resulting W2 accuracy, as summarized in Tables~\ref{tab:coeff_sweep} and \ref{tab:hp_sweep}.

\vspace{0.5em}
\noindent\textbf{Feature weights.}
We first vary each feature weight in Eq.~\ref{eq:outgoing}--\ref{eq:minor} to 0 (disabled) or 0.5 (emphasized) while keeping the others at the default of 0.25. Note that $\lambda_\text{BA}$, $\lambda_\text{PBD}$, and $\lambda_\text{KD}$ apply to all event categories, whereas $\lambda_\text{PBDS}$ and $\lambda_\text{OD}$ apply only to major (outgoing and incoming) and minor events, respectively.

\begin{table}[hb]
\centering
\footnotesize
\caption{Sensitivity of the W2 accuracy to the feature weights for each event category, where each weight is varied to 0 (disabled) or 0.5 (emphasized) while the others remain at the default ($\star$) of 0.25. Cells are shaded by the drop from the default as in Table~\ref{tab:coverage}.}
\vspace{-0.5em}
\label{tab:coeff_sweep}
\setlength{\tabcolsep}{1.3pt}
\begin{tabular}{l c | c C c C | C C}
\toprule
\textbf{Weight} & \textbf{Value} &
\makecell{\textbf{Open-play}\\\textbf{outgoing}} &
\makecell{\textbf{Set-piece}\\\textbf{outgoing}} &
\textbf{Incoming} & \textbf{Minor} &
\makecell{\textbf{Event}\\\textbf{start}} &
\makecell{\textbf{Event}\\\textbf{end}} \\
\midrule
Default$^\star$ & 0.25 & 97.8\% & 97.9\% & 90.2\% & 87.9\% & 96.5\% & 93.6\% \\
\midrule
\multirow{2}{*}{$\lambda_\text{BA}$}
 & 0.00 & 97.5\% & 97.5\% & \cy 88.5\% & \cd 70.1\% & \cy 94.9\% & \cy 92.6\% \\
 & 0.50 & 97.7\% & 97.9\% & \cy 86.8\% & \cy 86.1\% & 96.0\% & \cy 90.6\% \\
\midrule
\multirow{2}{*}{$\lambda_\text{PBD}$}
 & 0.00 & 97.7\% & 97.5\% & \cy 86.8\% & \cd 67.5\% & \cy 94.8\% & \cy 92.3\% \\
 & 0.50 & 97.2\% & 97.9\% & \cy 87.5\% & \cy 84.0\% & 95.6\% & 92.9\% \\
\midrule
\multirow{2}{*}{$\lambda_\text{KD}$}
 & 0.00 & \cl 91.7\% & \cy 94.7\% & \cl 76.6\% & \cd 71.4\% & \cl 89.4\% & \cl 79.6\% \\
 & 0.50 & 98.0\% & \cl 88.3\% & \cy 87.8\% & \cy 83.5\% & \cy 95.4\% & 94.1\% \\
\midrule
\multirow{2}{*}{$\lambda_\text{PBDS}$}
 & 0.00 & \cy 96.7\% & \cd 38.7\% & \cy 88.1\% & 87.4\% & \cl 90.7\% & 92.9\% \\
 & 0.50 & 97.1\% & \cy 95.7\% & 89.5\% & 87.9\% & 95.7\% & \cy 91.9\% \\
\midrule
\multirow{2}{*}{$\lambda_\text{OD}$}
 & 0.00 & 97.8\% & 97.9\% & 89.5\% & \cl 74.9\% & 95.6\% & 93.5\% \\
 & 0.50 & 97.8\% & 97.9\% & 90.5\% & \cl 79.7\% & 96.0\% & 93.6\% \\
\bottomrule
\end{tabular}
\vspace{-0.2em}
\end{table}

As shown in Table~\ref{tab:coeff_sweep}, disabling any feature ($\lambda = 0$) degrades accuracy in its corresponding categories, confirming that every feature contributes to the synchronization. Among them, $\lambda_{\text{KD}}$ has the broadest impact: removing it lowers the accuracy across all categories, as the kick distance is the key signal for distinguishing kicks, receptions, and minor touches. Meanwhile, disabling $\lambda_\text{PBDS}$ shows less effect on the other categories but collapses the set-piece accuracy to 38.7\%. This is because a set piece usually launches the ball from rest at the start of an episode, where the framework detects spurious candidates around the true frame. These candidates appear nearly identical to the true one in all features except the pre-slope, so the alignment often picks a wrong candidate without the PBDS score, missing the W2 tolerance.

Emphasizing a single feature ($\lambda = 0.5$) is also harmful, degrading the accuracy by overshadowing the other features. In particular, minor events are the most sensitive to these weight changes, as their inherently ambiguous motion patterns make the synchronization rely on the balance of all features. Across all categories, the default equal-weight setting yields the most consistent accuracy, justifying its use without further tuning.

\vspace{0.5em}
\noindent\textbf{Scoring function bounds.}
Next, we sweep the clipping bound of each scoring function in Eq.~\ref{eq:ball_accel}--\ref{eq:oppo_dist} around its default value. As shown in the upper block of Table~\ref{tab:hp_sweep}, the default value is the best or tied with the best for every bound, and the accuracy mostly varies within one percentage point across the swept ranges. The exceptions occur when a bound becomes too tight: clipping a feature at a small value saturates its score for most candidates, making large feature values indistinguishable from one another. Loosening the bounds instead mildly dilutes the resolution of the scores.

\vspace{0.5em}
\noindent\textbf{Alignment penalties.}
Finally, we vary the candidate gap penalty $g_c$, the event gap
penalty $g_e$, and the repeat penalty $r$ in Eq.~\ref{eq:ranw}. As reported in the lower block of Table~\ref{tab:hp_sweep}, the accuracy is entirely insensitive to the event gap penalty, supporting our choice of $g_e = 0$. In contrast, the candidate gap penalty degrades the accuracy in both directions, as a negative $g_c$ forces spurious matches by penalizing unmatched candidates, while a positive one rewards leaving true candidates unused. The repeat penalty behaves likewise: $r = 0$ over-merges consecutive events into a single frame, whereas $r \le -0.3$ blocks legitimate merges such as one-touch actions. Overall, the accuracy forms a wide plateau around the defaults, justifying our choice of hyperparameters.

\begin{table}[hb]
\caption{Sensitivity of the total W2 accuracy to the clipping bounds of the per-feature scoring functions and to the alignment penalties, where each parameter is varied while the others remain at their defaults ($\star$). Cells are shaded by the drop from the default as in Table~\ref{tab:coverage}.}
\label{tab:hp_sweep}
\footnotesize
\begin{tabular}{@{}l|ccccc@{}}
\toprule
\multicolumn{6}{@{}l}{\textbf{Scoring function bounds}} \\
\midrule
\quad BA bound (\si{m/s^2}) & 10 & 20 & 30$^\star$ & 40 & 50 \\
 & 94.7\% & 94.8\% & \textbf{95.2\%} & 94.9\% & 94.7\% \\
\midrule
\quad PBD/OD bound (\si{m}) & \cy 1 & 3$^\star$ & 5 & 7 & 9 \\
 & \cy 92.4\% & \textbf{95.2\%} & 95.1\% & 95.0\% & 94.7\% \\
\midrule
\quad KD bound (\si{m}) & \cl 1 & 3$^\star$ & 5 & 7 & 9 \\
 & \cl 89.1\% & \textbf{95.2\%} & \textbf{95.2\%} & 95.0\% & 94.7\% \\
\midrule
\quad PBDS bound (\si{m/s}) & \cy 1 & \cy 3 & 5 & 7$^\star$ & 9 \\
 & \cy 91.7\% & \cy 93.8\% & 94.7\% & \textbf{95.2\%} & \textbf{95.2\%} \\
\specialrule{1pt}{2pt}{3pt}
\multicolumn{6}{@{}l}{\textbf{Alignment penalties}} \\
\midrule
\quad Candidate gap $g_c$ & \cy $-0.3$ & $-0.1$ & $0^\star$ & \cy $+0.1$ & \cy $+0.3$ \\
 & \cy 93.9\% & 94.7\% & \textbf{95.2\%} & \cy 94.0\% & \cy 94.0\% \\
\midrule
\quad Event gap $g_e$ & $-0.3$ & $-0.1$ & $0^\star$ & $+0.1$ & $+0.3$ \\
 & \textbf{95.2\%} & \textbf{95.2\%} & \textbf{95.2\%} & \textbf{95.2\%} & \textbf{95.2\%} \\
\midrule
\quad Repeat penalty $r$ & \cy $0$ & $-0.1^\star$ & $-0.2$ & $-0.3$ & \cy $-0.4$ \\
 & \cy 94.0\% & \textbf{95.2\%} & 94.7\% & 94.4\% & \cy 93.9\% \\
\bottomrule
\end{tabular}
\vspace{-0.5em}
\end{table}

\subsection{Runtime Analysis}
\label{sec:runtime}
In this section, we measure the runtime of each synchronizer to verify that it does not introduce latency in practical analytics workflows. As reported in Table~\ref{tab:runtime}, all methods synchronize every match in under a minute, and our ELASTIC-NW takes under 40 seconds per match on average. The method of Biermann et al. is the fastest (1.4 seconds on average), as its inference reduces to a single pass of a lightweight classifier over the sliding windows once trained. DataBallPy also synchronizes a match only within 4.0 seconds on average, since it skips candidate frame selection and directly performs the NW alignment. However, this speed comes at the cost of coverage and accuracy, as the two baselines synchronize only a subset of event types. In contrast, our framework additionally detects ball receptions and prunes physically implausible frames through candidate frame selection, achieving substantially higher accuracy across all event categories (Table~\ref{tab:accuracy}). Moreover, since a 90-minute match is processed within a minute either way, such differences in runtime are practically irrelevant for post-match analysis.

\begin{table}[h]
\centering
\caption{Per-match runtime (in seconds) of each synchronizer on the three Sportec matches in the benchmark, along with the mean and standard deviation.}
\vspace{-0.5em}
\label{tab:runtime}
\setlength{\tabcolsep}{4pt}
\footnotesize
\begin{tabular}{l | r r r @{\hspace{6pt}} | r}
\toprule
\textbf{Method} & \multicolumn{1}{c}{\textbf{Match 1}} & \multicolumn{1}{c}{\textbf{Match 2}} & \multicolumn{1}{c|}{\textbf{Match 3}} & \multicolumn{1}{c}{\textbf{Mean $\pm$ Std}} \\
\midrule
ETSY            & 33.88 & 22.07 & 28.72 & $28.22 \pm 5.92$ \\
Biermann et al.      & 1.59  & 1.18  & 1.40  & $1.39 \pm 0.20$ \\
DataBallPy      & 4.66  & 3.43  & 3.92  & $4.00 \pm 0.62$ \\
ELASTIC-Greedy  & 37.12 & 30.06 & 29.70 & $32.29 \pm 4.19$ \\
ELASTIC-NW      & 43.67 & 34.94 & 34.27 & $37.62 \pm 5.25$ \\
\bottomrule
\end{tabular}
\end{table}

\section{Downstream Task Evaluation}
\label{sec:downstream}
Beyond direct synchronization accuracy, we examine how it affects downstream analytics tasks that leverage the synchronized event and tracking data. Many soccer analytics models take the spatial configuration of players at each event (e.g., \cite{anzer2021goal, anzer2022expected, fernandez2020soccermap, fernandez2021framework, kim2026better, spearman2017physicsbased, robberechts2023unxpass, wang2024tacticai}) as input, so an inaccurate event timestamp places the players at the wrong positions and feeds a distorted snapshot to the model. We therefore expect that more accurate synchronization leads to more reliable training and evaluation of such models. To demonstrate this, we measure the performance of two representative prediction tasks based on Graph Neural Networks (GNNs) when their input snapshots are synchronized by different methods.

\begin{table}[t!]
\centering
\caption{Downstream task performance under different synchronization methods. Arrows indicate whether a higher ($\uparrow$) or lower ($\downarrow$) value is better.}
\label{tab:downstream}
\setlength{\tabcolsep}{4pt}
\footnotesize
\begin{tabular}{l | r r r | r r r}
\toprule
 & \multicolumn{3}{c|}{\textbf{Next action prediction}} & \multicolumn{3}{c}{\textbf{Pass success prediction}} \\
\cmidrule(lr){2-4} \cmidrule(lr){5-7}
\textbf{Method} & \multicolumn{1}{c}{\textbf{Acc.}\,$\uparrow$} & \multicolumn{1}{c}{\textbf{CE}\,$\downarrow$} & \multicolumn{1}{c|}{\textbf{MRR}\,$\uparrow$} & \multicolumn{1}{c}{\textbf{F1}\,$\uparrow$} & \multicolumn{1}{c}{\textbf{AUC}\,$\uparrow$} & \multicolumn{1}{c}{\textbf{Brier}\,$\downarrow$} \\
\midrule
Unsynced       & 0.5796 & 1.2148 & 0.7354 & 0.9031 & 0.9028 & 0.1013 \\
ELASTIC-Greedy & 0.6759 & 0.9114 & 0.8017 & 0.9072 & 0.9103 & 0.0966 \\
ELASTIC-NW     & \textbf{0.6863} & \textbf{0.8773} & \textbf{0.8087} & \textbf{0.9155} & \textbf{0.9172} & \textbf{0.0899} \\
\bottomrule
\end{tabular}
\vspace{-0.5em}
\end{table}

\begin{figure}[t!]
    \centering
    \begin{subfigure}[t]{\columnwidth}
        \centering
        \includegraphics[width=0.9\columnwidth]{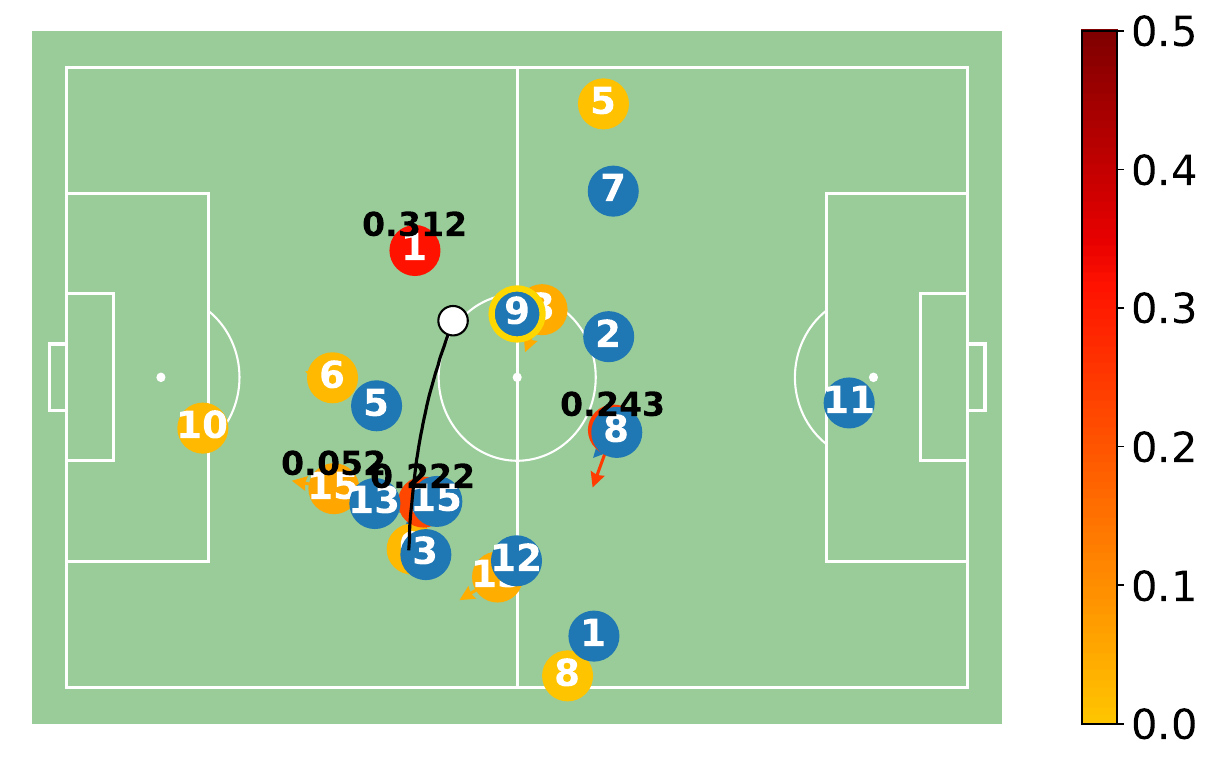}
        \vspace{-1em}
        \caption{Next action probabilities}
        \label{fig:select_probs}
    \end{subfigure}
    \begin{subfigure}[t]{\columnwidth}
        \centering
        \includegraphics[width=0.9\columnwidth]{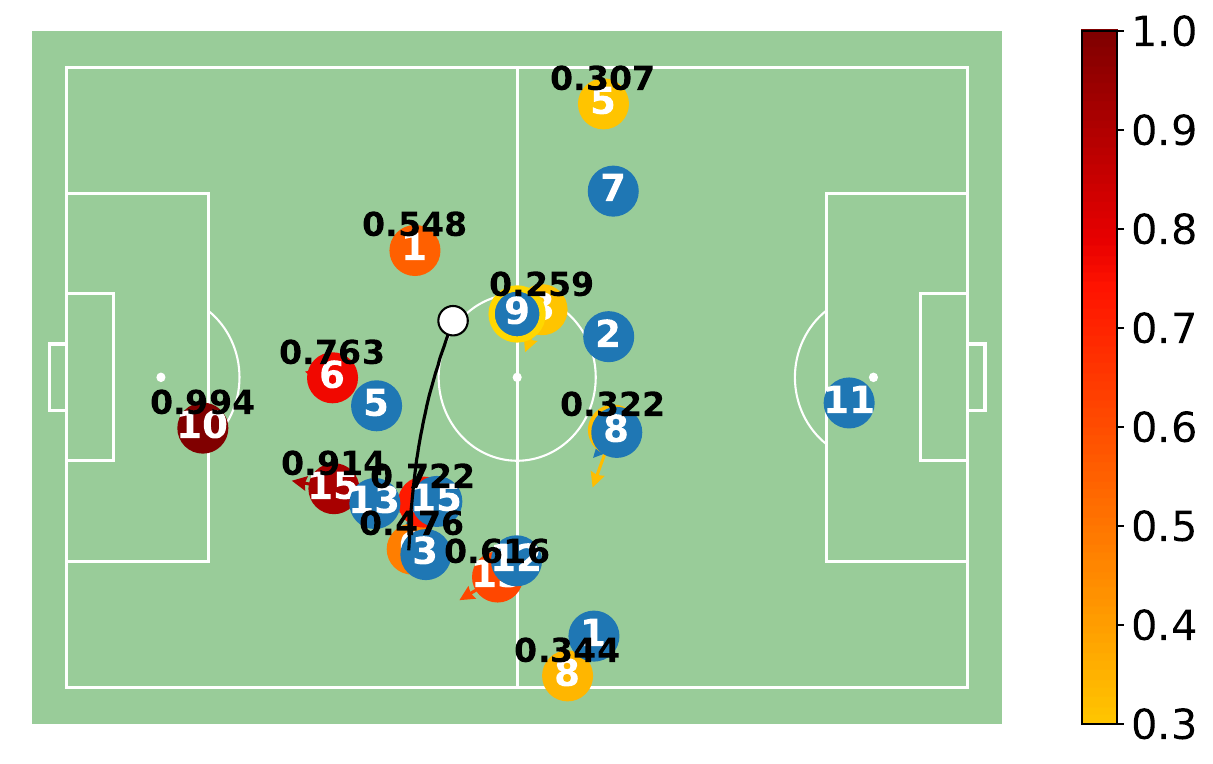}
        \vspace{-1em}
        \caption{Pass success probabilities}
        \label{fig:success_probs}
    \end{subfigure}
    \caption{Per-player outputs of the two downstream models for a sample pass, trained on data synchronized by ELASTIC-NW. Each value, shown both as text and by the circle's color, indicates (a) the probability that the ball possessor selects each teammate as the target and (b) the probability that a pass to the teammate would succeed.}
    \label{fig:downstream}
\end{figure}

\subsection{Tasks and Setup}
\label{sec:downstream_tasks}
We consider the following two tasks that take the player and ball configuration at each event as a graph input and predict an attacking outcome. For both tasks, we follow the GNN architecture of Kim et al.~\cite{kim2026better}, representing each game state as a fully connected graph of players and two goals and producing node embeddings with Graph Attention Network (GAT)~\cite{velickovic2018graph} layers.
\begin{itemize}
    \item \textbf{Next action prediction} estimates which action the ball possessor will attempt next, namely a pass to one of the teammates, a dribble (treated as a pass to oneself), or a shot toward the opponent's goal. We model it as a node selection task over the possessor's teammates and the target goal, applying a softmax over their node embeddings to produce a probability distribution that sums to one. Since this task is a multi-class classification problem, we use accuracy, cross-entropy (CE), and mean reciprocal rank (MRR) of the true node as performance metrics in Table~\ref{tab:downstream}.
    \item \textbf{Pass success prediction} estimates whether a pass to each teammate would succeed if attempted, applying a sigmoid to each node embedding to produce an independent success probability. Since this task is a binary classification task (whether the pass would succeed or fail), we use F1 score, AUC, and Brier score as performance metrics in Table~\ref{tab:downstream}.
\end{itemize}
Fig.~\ref{fig:downstream} illustrates the per-player outputs of the two models for a pass. Both tasks play a central role in understanding game situations~\cite{anzer2022expected, fernandez2021framework, spearman2018beyond} and evaluating players' performance~\cite{everett2025evaluating, kim2026better, robberechts2023unxpass, stockl2021making}.

Since the Sportec Open DFL Dataset contains only seven matches, we use a larger proprietary tracking dataset from the Dutch Eredivisie to train and evaluate the models, with 200 matches from the 2023--24 season for training, 50 matches from the same season for validation, and 157 matches from the 2024--25 season for testing. We train and evaluate each task independently on three versions of the data: the unsynchronized event data, and the data synchronized by ELASTIC-Greedy and ELASTIC-NW, respectively.

\subsection{Results}
\label{sec:downstream_results}
As shown in Table~\ref{tab:downstream}, data synchronization consistently improves downstream performance. Compared to the unsynchronized data, even ELASTIC-Greedy raises accuracy on both tasks (especially the next-action prediction accuracy from 0.5796 to 0.6759), confirming that accurate synchronization is an essential preprocessing step for soccer analytics. Moreover, ELASTIC-NW further outperforms ELASTIC-Greedy across all metrics, demonstrating that the gains from its global alignment translate into measurable downstream benefits. This indirect comparison on a much larger set of matches thus corroborates the direct evaluation results reported in Section~\ref{sec:experiments}, confirming the benefit of ELASTIC-NW to practical analytics.

\section{Conclusion}
\label{sec:conclusion}
This paper revisits event-tracking synchronization in soccer from two angles that prior work has left open: removing the dependence on noisy human-annotated event locations, and recovering the ball receptions that mark when each event ends. ELASTIC addresses both by reasoning purely over player and ball trajectories, and our re-annotated public benchmark shows that this design not only achieves the most accurate synchronization to date but also uniquely recovers event endings. Crucially, our downstream evaluation reframes synchronization as more than a preprocessing detail: the quality of alignment directly shapes the reliability of the analytics built on top of it. We hope that both our method and the released benchmark lower the barrier to sports data analytics.

\section*{Acknowledgments}
This work was supported by the National Research Foundation of Korea (NRF) grant funded by the Korea government (MSIT) (RS-2024-00335098, RS-2024-00406985), and by the National Research Foundation of Korea (NRF) funded by Ministry of Science and ICT (RS-2022-NR068758).

\section*{GenAI Usage Disclosure}
We used generative AI tools in a limited and supportive manner during the preparation of this manuscript. Specifically, we used Claude to improve the clarity and readability of the writing through phrasing and grammar refinement. In addition, we used Claude Code to support data preprocessing, visualization, and debugging. All core research contributions, including the main ideas, methodology design, experiments, and analytical insights, were developed and validated entirely by the authors.

\balance
\bibliography{main}
\bibliographystyle{ACM-Reference-Format}

\end{document}